\documentclass{aa}
\usepackage{psfig}
\newcommand{\diver}{\mathop{\rm div}\nolimits}
\newcommand{\grad}{\mathop{\rm grad}\nolimits}

\begin{document}

\thesaurus{ 02.13.2;08.05.3;08.19.4}

\title{Nonstationary magnetorotational processes in a rotating
       magnetized cloud}

\author{
        N.V.Ardeljan \inst{1}\and
        G.S.Bisnovatyi-Kogan \inst{2} \and
        S.G.Moiseenko \inst{2}
       }

\offprints{S.G. Moiseenko, \\ moiseenko@mx.iki.rssi.ru}

\institute{
           Department of Computational Mathematics
           and Cybernetics, Moscow State University,
           Vorobjevy Gory, \\
           Moscow B-234 119899, Russia, ardel@redsun.cs.msu.su
           \and
           Space Research Institute, Profsoyuznaya
           84/32, Moscow 117810, Russia \\
           gkogan@mx.iki.rssi.ru, moiseenko@mx.iki.rssi.ru
          }

\date{Received ; accepted }

\authorrunning{Ardeljan, Bisnovatyi-Kogan, Moiseenko}
\titlerunning{magnetorotational processes in cloud}

\maketitle

\begin{abstract}

We perform 2D numerical simulations of a magnetorotational
explosion of a rotating magnetized gas cloud. We found
that amplification of a toroidal magnetic field due to the
differential rotation leads to a transformation of the part of
the rotational energy of the cloud to the radial kinetic energy.
Simulations have been made for 3 initial values of $\xi$
(the relation
of magnetic energy to the gravitational energy of the cloud):
$\xi =10^{-2},\>10^{-4},\>10^{-6}$. Part of the matter -
$\sim 7\%$ of the mass of the cloud ($\sim 3.3\%$ of the final
gravitational energy of the
cloud) - gets radial kinetic energy which is larger than its
potential
energy and can be thrown away to the infinity. It carries about
30\% of the initial angular momentum of the cloud. This effect is
important for angular momentum loss in the processes of stellar
formation, and for the magnetorotational mechanism of explosion
suggested for supernovae. Simulations have been made on the basis
of the Lagrangian 2D numerical implicit scheme on a triangular
grid with grid reconstruction.

\end{abstract}



\section{Introduction}


Magnetic field plays an important role in life of a star,
especially at its birth and its death. A birth of a star in a
galactic disk during the collapse of an interstellar cloud
is complicated by large angular momentum of matter which will
prevent formation of the star if it is not lost. The most
realistic mechanism of the angular momentum loss is connected
with a magnetic field which, twisting during differential
rotation, leads to a flux of angular momentum outside, permiting
a collapse of the main body (\cite{bkruzsun}, Ardeljan et al. 1996b).

In the late stages of evolution of massive stars, loss of
hydrodynamical stability initiates a collapse, which is finished
by supernovae explosions after formation of a stable neutron
star. Investigation of supernovae explosions in different
nonmagnetized models revealed serious problems in transformation
of the gravitational energy into the energy of explosion. In
this situation magnetorotational mechanism of supernovae
explosion, suggested by Bisnovatyi-Kogan (1970) could be an
explanation of this phenomena.

Magnetorotational phenomena in stellar envelopes are considered
to be a main mechanism  of an angular momentum loss from stars,
and magnetized solar wind is a reason for a very slow solar
rotation.

2D calculations of collapse of a rotating magnetized star with
a somewhat unrealistic magnetic field configuration have been
performed by Le Blank \& Wilson (1970) and Ohnishi (1983). The
geometry of the outburst which they obtained, was different,
but the energy of the outburst was substantial in both cases.

We investigate here magnetorotational phenomena in the
collapsing rotating magnetized gas cloud initially with uniform
density and angular momentum, which may be considered as a model
of star formation. After magnetorotational explosion a
quasistationary, almost uniformly slow rotating configuration is
formed.
The loss of angular momentum is connected with an outburst of
the matter, carrying an energy which is $\sim$ 3.3\% of
the final gravitational energy of the cloud. Our simulations
show that amounts of ejected mass and energy are weakly
dependent on the initial value of magnetic energy (parameter
$\xi$).
The main
difference in the results of 2D simulations for the different
initial values of $\xi$ is in the duration of the process.
This conclusion confirms results of 1D simulations
(\cite{bkpopsam}, \cite{ardbkpop}), with the time of the process
approximately proportional to the $\xi^{-1/2}$.
Similar processes related to a different initial model and
equation of state are expected to act during supernovae
explosions.

Magnetorotational explosion of a rotating magnetized gas cloud
in a 2D approach has been investigated earlier by
\cite{ardmsgass} for a divergency-free, but not force-free,
magnetic field configuration. Magnetic forces have not been
balanced by a pressure distribution and gravitational forces.
The effect of ejection of the part of the mass of the cloud has
been found there, but use of an unbalanced initial field leads to
artificial effects in a MHD flow and does not allow us to extend
calculations as far as is needed.

In this paper an initial configuration with magnetic forces
balanced by pressure and gravitational forces distribution has
been constructed, allowing us to make calculations for
different sets of initial parameters and to extend them to the
larger times.


\section{Basic equations}
\label{basiceq}


Consider a set of magnetohydrodynamical equations with
self\-gra\-vi\-ta\-tion and with infinite conductivity
(Lan\-dau \& Lif\-shitz, 1984, \cite{bkbook}):
\begin{eqnarray}
\frac{{\rm d} {\bf x}} {{\rm d} t} = {\bf u}, \quad
\frac{{\rm d} \rho} {{\rm d} t} +
\rho \diver {\bf u} = 0,  \nonumber
\end{eqnarray}
\begin{eqnarray}
\rho \frac{{\rm d} {\bf u}}{{\rm d} t} =-{\rm grad}
\left(p+\frac{{\bf H} \cdot {\bf H}}{8\pi}\right) +
\frac {\diver({\bf H} \otimes {\bf H})}{4\pi} -
\rho  \grad \Phi,
\nonumber
\end{eqnarray}
\begin{eqnarray}
\rho \frac{{\rm d}}{{\rm d} t} \left(\frac{{\bf H}}{\rho}\right)
={\bf H} \cdot \nabla {\bf u},\quad
\Delta \Phi=4 \pi G \rho,
\label{magmain}
\end{eqnarray}
\begin{eqnarray}
\rho \frac{{\rm d} \varepsilon}{{\rm d} t} +p \diver {\bf u}=0, \quad
\frac {1}{\rho}=\frac{T {\Re}}{p}, \quad
\varepsilon=\frac{T {\Re}}{\gamma-1}, \nonumber
\end{eqnarray}
where $\frac {\rm d} {{\rm d} t} = \frac {\partial} {
\partial t} + {\bf u} \cdot \nabla$ is the total time
derivative, ${\bf x} = (r,\varphi , z)$, ${\bf u}$ is velocity
vector, $\rho$ is density, $p$ is pressure,  ${\bf
H}=(H_r,\> H_\varphi,\> H_z)$ is magnetic field vector, $\Phi$ is
gravitational potential, $\varepsilon$ is internal energy, $G$ is
gravitational constant, $\Re$ is universal gas constant,
$\gamma$ is adiabatic index, ${\bf H} \otimes {\bf H}$ is tensor
of rank 2.

Axial symmetry ($\frac \partial {\partial
\varphi}$) and symmetry to the equatorial plane ($z=0$) are
assumed.

The problem is solved in the restricted domain. Outside
the domain, the density of the matter is zero, but poloidal
components of magnetic field $H_r,\> H_z$ can be non-zero.

To write the set of equations in dimensionless form we choose
the following scale values:
\begin{eqnarray}
\rho_0=1.492 \cdot 10^{-17}  {\rm g/cm^3},\>
r_0=z_0=x_0=3.81 \cdot 10^{16}{\rm cm},
\nonumber
\end{eqnarray}
\begin{eqnarray}
r={\tilde r}x_0, \> z={\tilde z}x_0, \> u={\tilde u}u_0,\>
u_0=\sqrt{4\pi G \rho_0 x_0^2},
\nonumber
\end{eqnarray}
\begin{eqnarray}
p={\tilde p}p_0, \> \varepsilon={\tilde \varepsilon}
\varepsilon_0, \> T={\tilde T}T_0, \> \Phi={\tilde \Phi}\Phi_0,
\> \Phi_0=4\pi G \rho_0 x_0^2,
\nonumber
\end{eqnarray}
\begin{eqnarray}
t_0=\frac {x_0} {u_0},\>
p_0=\rho_0 u_0^2=\rho_0
x_0^2 t_0^{-2}, \>  T_0= \frac {u_0^2}{\Re},
\nonumber
\end{eqnarray}
\begin{eqnarray}
\varepsilon_0=u_0^2=x_0^2t_0^{-2}, \>
H_0=\sqrt {p_0}=x_0t_0^{-1} \rho_0^{1/2}.
\nonumber
\end{eqnarray}
Here the values with index zero are the scale factors and the
functions under a tilde are dimensionless functions.
The set of equations (\ref{magmain}) can be
written in the following nondimentional form (the tilde being omitted):
\begin{eqnarray}
\frac{{\rm d} {\bf x}} {{\rm d} t} = {\bf u}, \quad
\frac{{\rm d} \rho} {{\rm d} t} +
\rho \diver {\bf u} = 0,  \nonumber
\end{eqnarray}
\begin{eqnarray}
\rho \frac{{\rm d} {\bf u}}{{\rm d} t} =-{\rm grad}
\left(p+\frac{{\bf H} \cdot {\bf H}}{8\pi}\right) +
\frac {\diver({\bf H} \otimes {\bf H})}{4\pi} -
\rho  \grad \Phi,
\nonumber
\end{eqnarray}
\begin{eqnarray}
\rho \frac{{\rm d}}{{\rm d} t} \left(\frac{{\bf H}}{\rho}\right)
={\bf H} \cdot \nabla {\bf u}, \quad
\Delta \Phi= \rho,
\label{magdim}
\end{eqnarray}
\begin{eqnarray}
\rho \frac{{\rm d} \varepsilon}{{\rm d} t} +p \diver {\bf u}=0, \quad
\frac {1}{\rho}=\frac{T}{p}, \quad
\varepsilon=\frac{T}{\gamma-1}. \nonumber
\end{eqnarray}

Taking into account symmetry assumptions
($ \frac \partial {\partial \varphi} = 0$),
the divergency of the
tensor ${\bf H} \otimes {\bf H}$ can be presented in the
following form:
$$
{\rm div}({\bf H} \otimes {\bf H})=
\left(\begin{array}{l}
\frac {1}{r} \frac {\partial(rH_rH_r)}{\partial r} +
\frac {\partial(H_zH_r)} {\partial z}-
\frac {1}{r} H_\varphi H_\varphi \\
\frac {1}{r} \frac {\partial(rH_rH_\varphi)}{\partial r} +
\frac {\partial(H_zH_\varphi)} {\partial z}+
\frac {1}{r} H_\varphi H_r \\
\frac {1}{r} \frac {\partial(rH_rH_z)}{\partial r} +
\frac {\partial(H_zH_z)} {\partial z}
\end{array}
\right).
$$


\section{Collapse of a rotating magnetized gas cloud}



\subsection{Formulation of the problem}


Consider a magnetized rotating gas cloud which is described by
the set of equations (\ref{magmain}).
All graphs and figures below are in a nondimensional form.
At the initial moment ($t=0$)
the cloud is a rigidly rotating uniform gas sphere (Fig. \ref{grid0})
with the following parameters:
\begin{figure}
\centerline{\psfig{figure=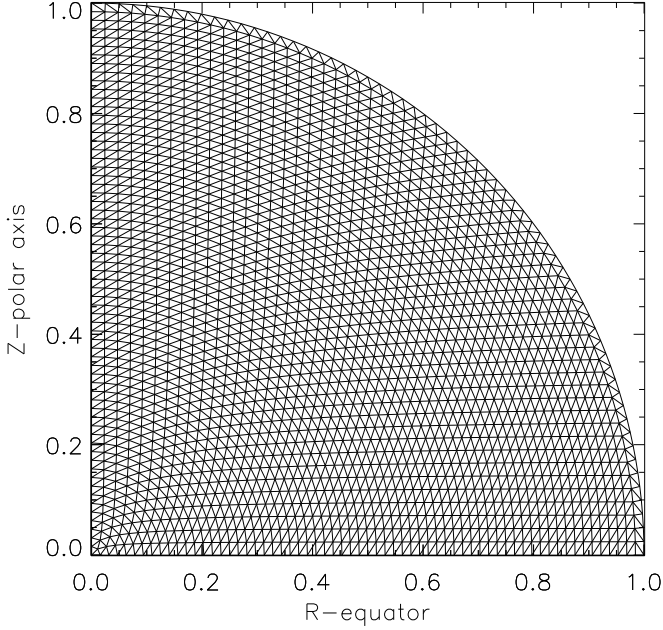,height=9.5cm,width=9.5cm}}
\caption{Grid at initial time $t=0$.}
\label{grid0}
\end{figure}
\begin{eqnarray}
r=3.81 \cdot 10^{16}{\rm cm}, \> \rho=1.492 \cdot 10^{-17}
{\rm g/cm}^3,\> \eta= 5/3,\nonumber
\end{eqnarray}
\begin{eqnarray}
M=1.73M_\odot=3.457\cdot 10^{33} {\rm g},\>
 u^r=u^z=0, \label{inicond}
\end{eqnarray}

\begin{eqnarray}
\beta_{r0}=\frac {E_{\rm rot0}}{\vert E_{\rm gr0}\vert }
=0.04,\label{alpha}\\
\beta_{i0}=\frac {E_{\rm in0}}{\vert E_{\rm
gr0}\vert } =0.01,\label{betta}\\
\xi=\frac {E_{\rm
mag1}}{\vert E_{\rm gr1}\vert } = 10^{-2},\>10^{-4},10^{-6}.
\label{gamma}
\end{eqnarray}
Here the subscript "0" corresponds to
the initial moment ($t=0$), subscript "1" to the moment of the
beginning of the evolution of magnetic field ($t=t_1>0$).

The assumptions about the symmetry of the problem are the same as in
Section \ref{basiceq}.

The process can be divided into the following three
qualitatively different stages. First is a pretty short
hydrodynamical collapse stage. At this stage the influence of the
magnetic field on the process of the collapse of the cloud can
be neglected, because the initial poloidal magnetic field is
weak. In the short time of collapse the toroidal
component of the magnetic field, which appears to be due to the
arising differential rotation is also weak at this initial
stage. The second stage is the stage of a rather ''long''
twisting of magnetic field due to the differential rotation
of the cloud. The final third stage starts with the appearance of
a compression wave, moving from the inner parts of the cloud to its
periphery along a steeply decreasing density background. Soon after
its appearance, it transforms into the MHD shock wave, which can
push out a light envelope of the protostar. Similar ejection
following formation of a rapidly rotating neutron star and a
differentially rotating envelope, can be interpreted as supernova
explosion.

Simulations for different $\xi$ consist of the calculation
of oscillations of the cloud until formation of the
differentially rotating equilibrium (without magnetic field), and
subsequent inclusion and twisting of the magnetic field.


\subsection{Initial magnetic field configuration}


The initial magnetic field must satisfy the condition of absence of
magnetic charges ${\rm div} {\bf H} = 0$. It also has to
correspond to the boundary conditions of the problem.

The best choice would be dipole or quadrupole. While they
satisfy initial and boundary conditions, they have singularities
in the origin of coordinates ($r=0,\> z=0$). Using such
magnetic fields in numerical simulations can lead to loss of
accuracy of calculations.

To define the initial magnetic field we use the following method.
We defined toroidal current $j_\varphi$ in the central part of
the core of the collapsed cloud by a formula:
\begin{eqnarray}
j_\varphi & = & j_\varphi^u+j_\varphi^d,\label{inicur}\\
j_\varphi^u & = & \Big[{\rm sin}\left(\pi {r\over 0.3} -
              {\pi \over 2}\right)+1\Big]
          \Big[{\rm sin}\left(\pi {z\over 0.3} -
              {\pi \over 2}\right)+1\Big]\nonumber\\
         &&\times\Big[1-\left({r\over 0.3}\right)^2
            -\left({z\over 0.3}\right)^2\Big],\nonumber\\
         && \> {\rm at}\> r^2+z^2< 0.3^2, \> z>0, \nonumber\\
j_\varphi^d&=&-j_\varphi^u \> {\rm at}\>
    r^2+z^2< 0.3^2, \> z<0.\nonumber
\end{eqnarray}

\begin{figure}
\vspace{-0.9cm}
\centerline{\psfig{figure=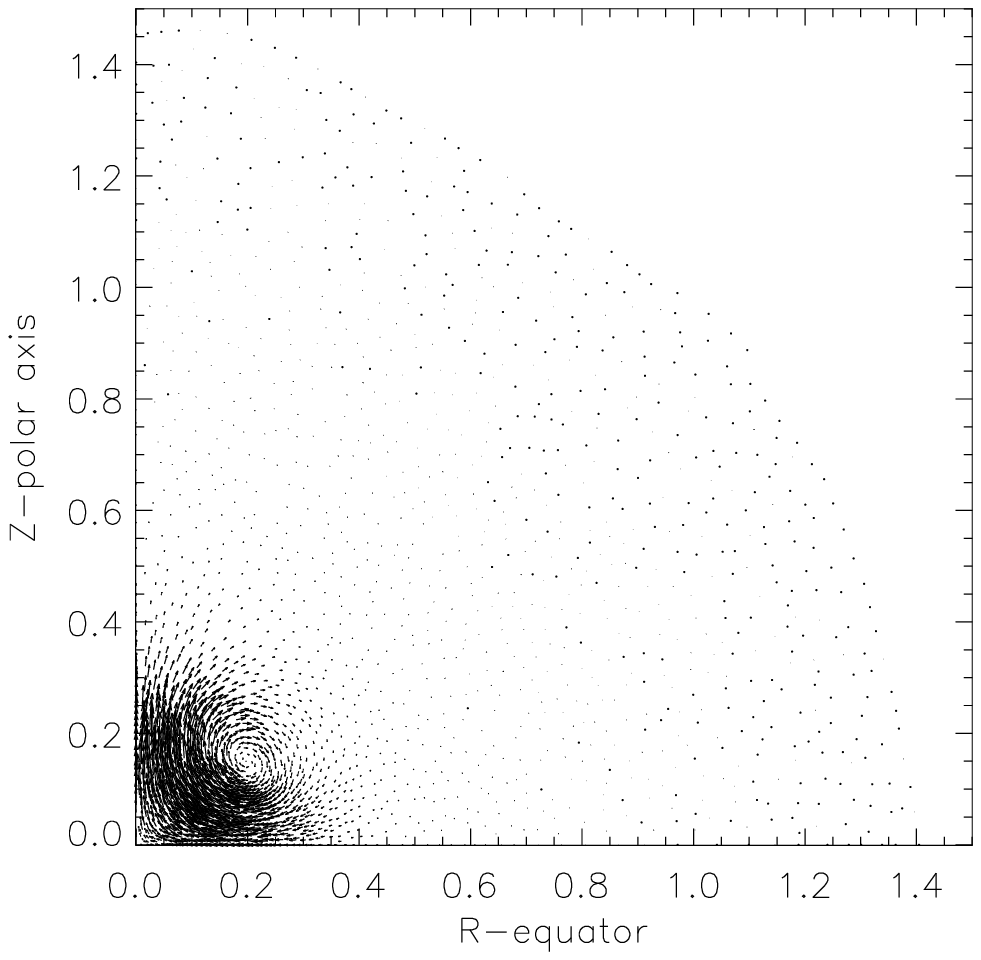,height=8.5cm,width=8.5cm}}
\vspace{-8.5cm}
\centerline{\psfig{figure=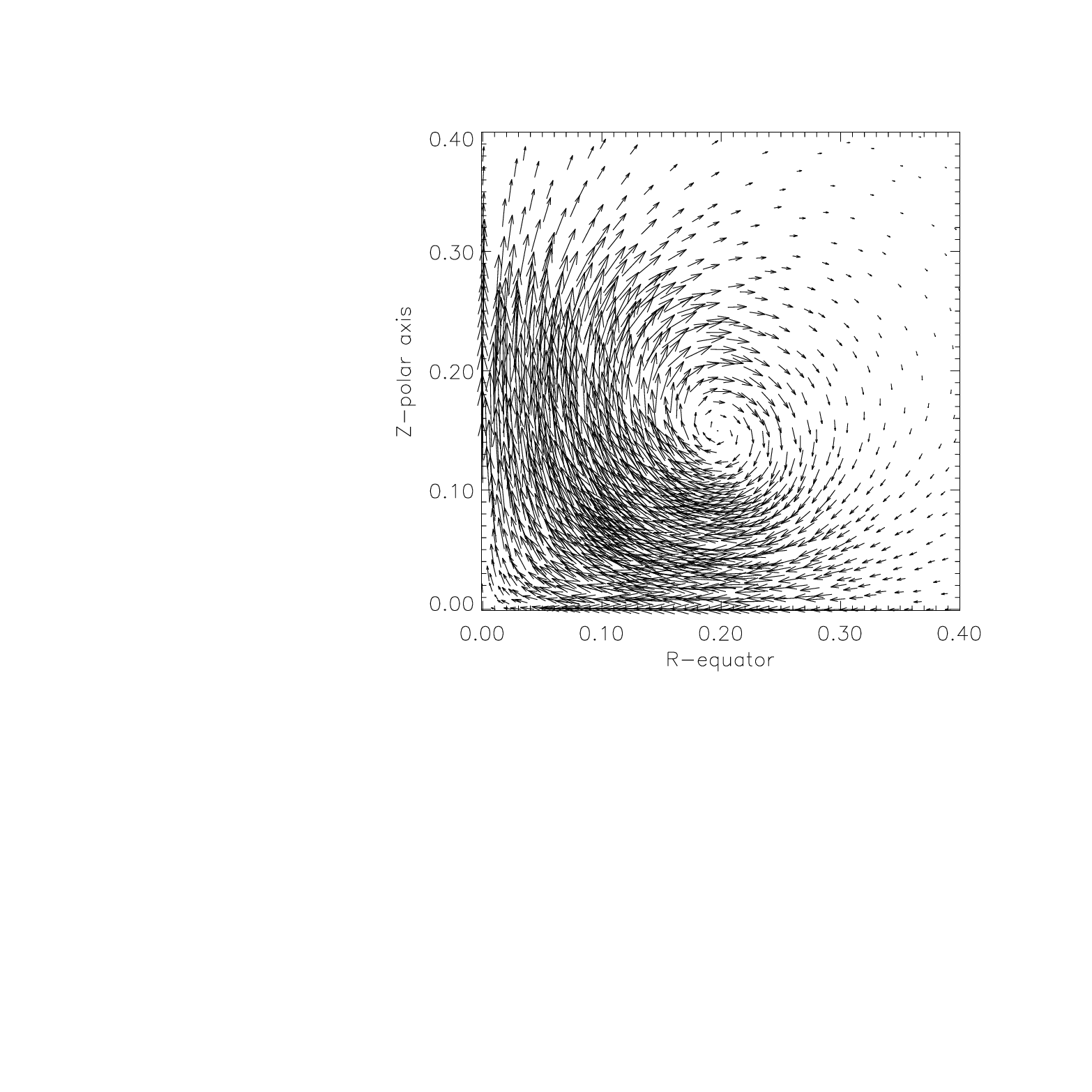,height=8.5cm,width=8.5cm}}
\caption{Initial magnetic field, produced by current (\ref{inicur}).
Upper right part of the figure is the magnetic field in the
central part of the cloud.}
\label{inimag}
\end{figure}

After getting the defferentially rotating stationary solution
for nonmagnetized cloud we use Bio-Savara law
(\ref{biosavara}),(\ref{biosavararas}),(\ref{biosavr}) for
calculation of the poloidal components of the magnetic field
$H_{r0},\> H_{z0}$ (Fig. \ref{inimag}). This magnetic field is
divergency free, but it is not force-free and should be
balanced at the initial moment. Then we use the following
method:  we "turn on" the poloidal magnetic field
$H_{r0},\>H_{z0}$, but "switch off" the equation for the
evolution of the toroidal component $H_\varphi$ in
(\ref{magmain}). Actually it means that we define $H_\varphi
\equiv 0,\> {{\rm d}H_\varphi \over {\rm d} t} \equiv 0$.  From
the physical point of view it means that we allow magnetic field
lines to slip through the matter of the cloud in the toroidal
direction.  After "turning on " such a field, we let the cloud
come to the steady state, where magnetic forces connected
with the purely poloidal field are balanced by other forces.

The calculated balanced configuration has the magnetic field of
quadrupole-like symmetry.  For  testing we run our code with
this purely poloidal field for a large number of time steps ($\sim
10^3$), during which the parameters of the cloud did not change.


\section{Results}

We describe results of simulations of the problem for initial
$\xi(t_1)=10^{-2}$. Results of simulations for
$\xi(t_1)=10^{-4},\> 10^{-6}$ are qualitatively similar to the
first case. The amounts of the ejected mass and energy are the
same, $\sim 7\%$ of mass and $\sim 3.3\%$ of energy. The main
difference betveen these three variants is the duration of
the process.
The lower the initial magnetic energy, the longer the
evolution time up to the magnetorotational explosion. This
feature of the results of the simulations in a 2D case is similar
to the results of 1D simulations of the magnetorotational
mechanism for supernovae (Ardeljan, et.al. 1979).

\begin{figure}
\vspace{-0.8cm}
\centerline{\psfig{figure=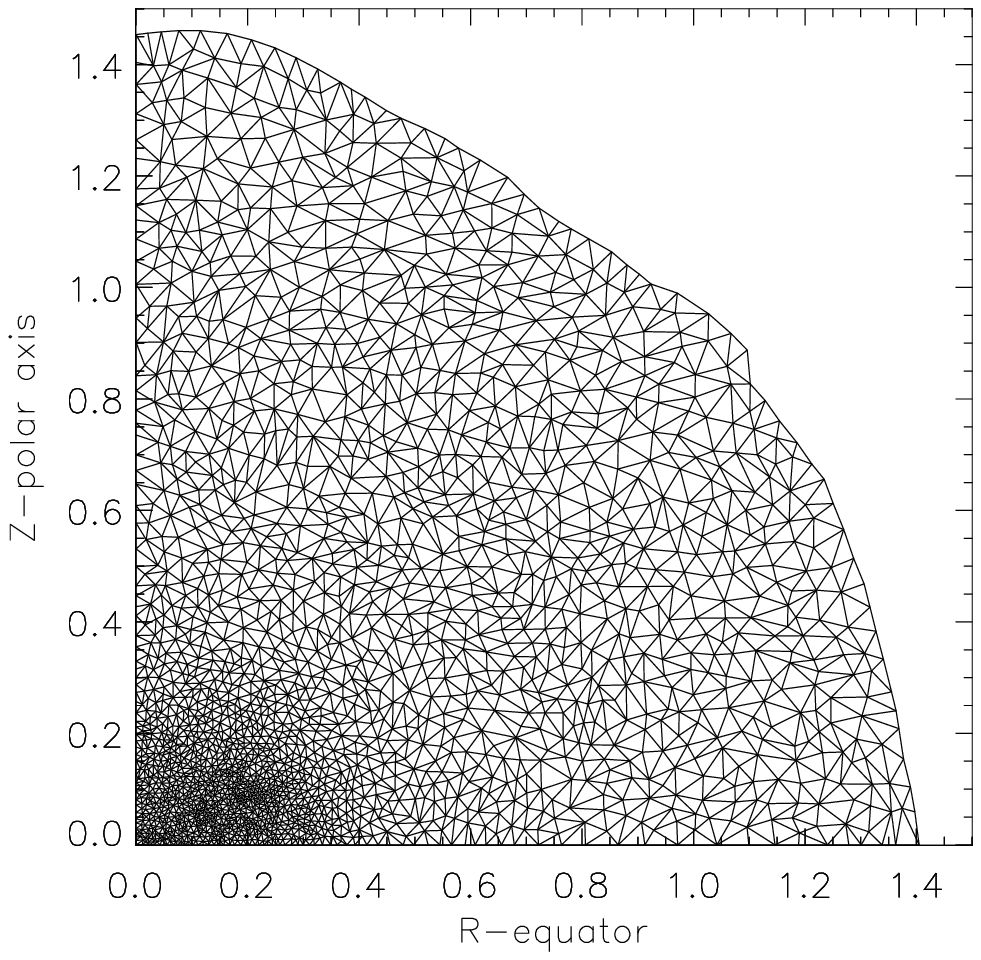,height=8.5cm,width=8.5cm}}
\caption{Lagrangian grid at $t_1=18.45$.}
\label{grid1}
\end{figure}
\begin{figure}
\vspace{-0.8cm}
\centerline{\psfig{figure=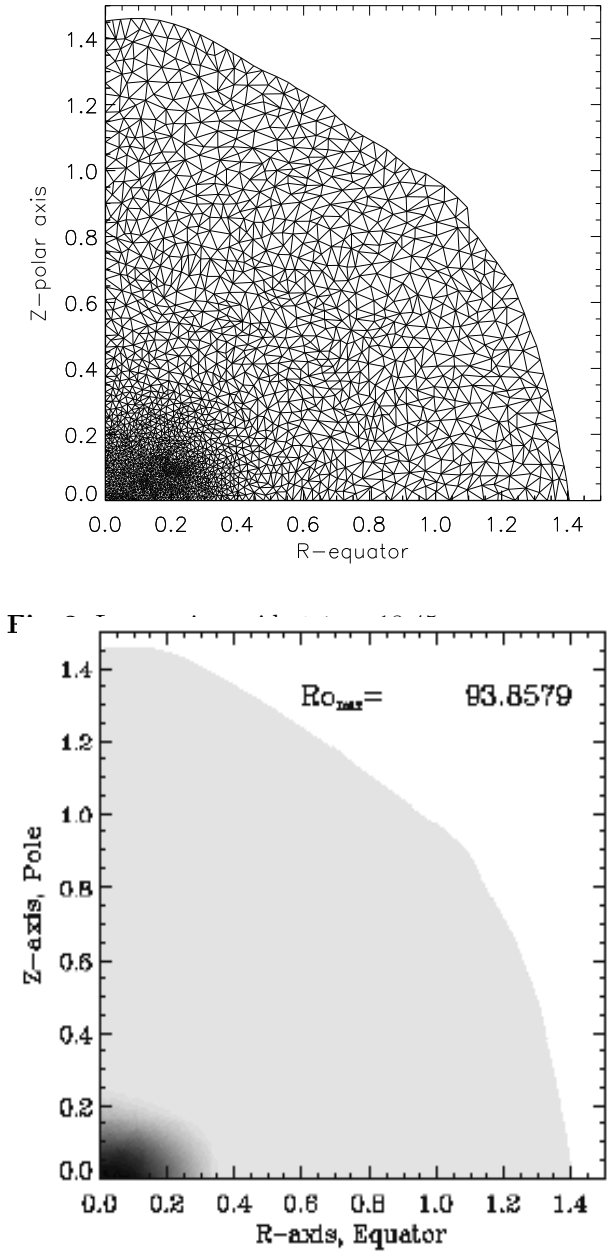,height=8.5cm,width=8.5cm}}
\caption{Density field at $t_1=18.45$.}
\label{rour1}
\end{figure}
\begin{figure}
\centerline{\psfig{figure=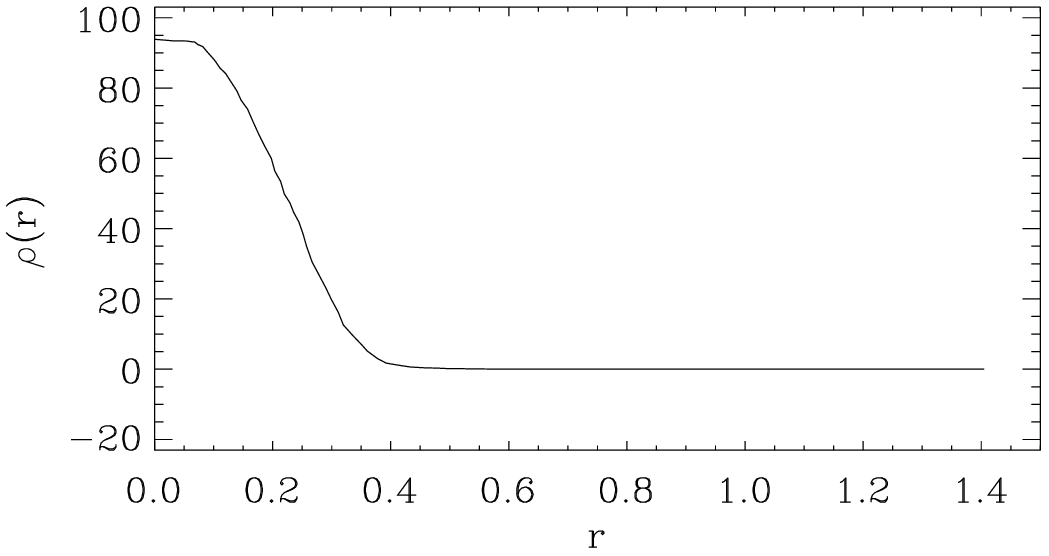,height=4.6cm,width=8.5cm}}
\caption{Density distribution along $r$ axis at $t_1=18.45$.}
\label{ror0}
\end{figure}
\begin{figure}
\vspace{-0.8cm}
\centerline{\psfig{figure=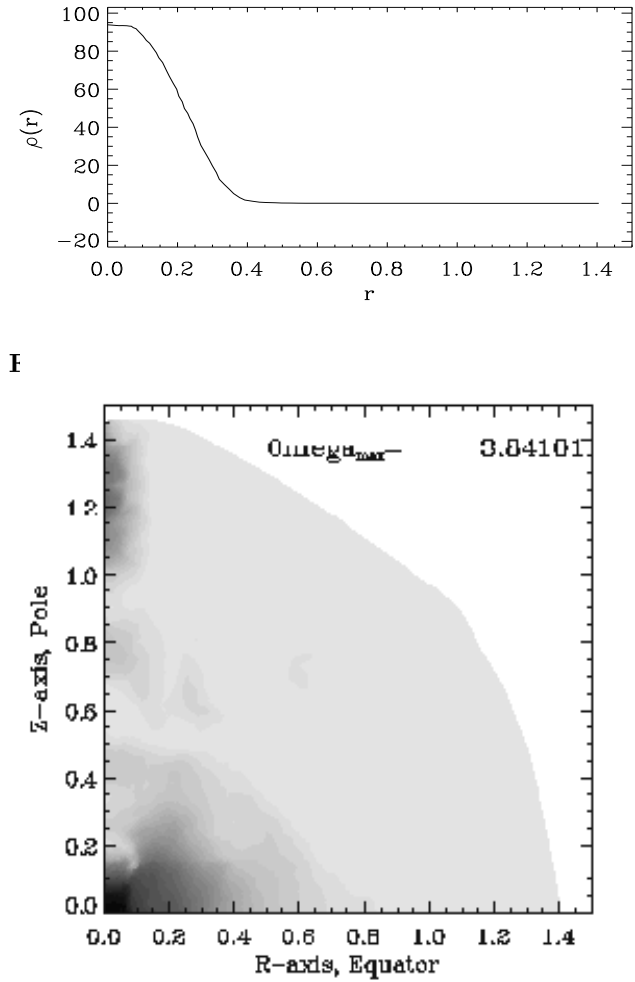,height=8.5cm,width=8.5cm}}
\caption{Angular velocity levels at $t_1=18.45$.}
\label{angvel1}
\end{figure}
\begin{figure}
\centerline{\psfig{figure=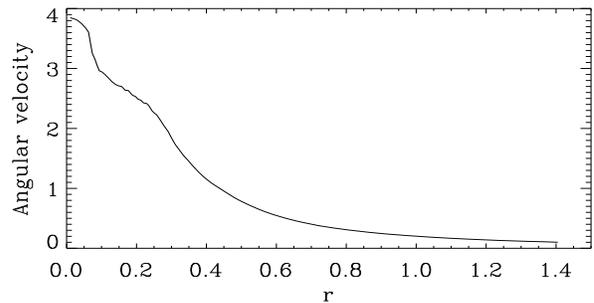,height=4.6cm,width=8.5cm}}
\caption{Angular velocity distribution along $r$ axis at $t_1=18.45$.}
\label{omegar0}
\end{figure}

We start calculations ($t=0$) with the set of initial parameters
(\ref{inicond})--(\ref{betta}) (without magnetic field at the
beginning).  Our triangular grid at $t=0$ consists of 2200 knots
and 4400 cells. During the calculations, the number of knots and
cells deviates no more than 5\% from its initial values due
to specially developed procedure (Ardeljan, et.al. 1996a).
After a number of oscillations the cloud at $t_1=18.45$ comes to
the steady differentially rotating configuration. Lagrangian
grid, density field and angular velocity levels are shown in
Figs.  \ref{grid1}, \ref{rour1}, \ref{angvel1}. Distribution of
the density and angular velocity for the $t_1=18.45$ along
$r$-axis are given in Figs. \ref{ror0}, \ref{omegar0}.

At $t_1$ the cloud consists of a rapidly rotating dense core and a
slowly rotating prolate envelope (Figs. \ref{ror0}, \ref{omegar0}).
The radial part of kinetic
energy of the cloud is less than $1.5\%$ of its gravitational
energy. The cloud can stay an infinite time in such differentially
rotating stationary condition (if we neglect dissipation
processes). However inclusion of even a weak magnetic field
leads to an essential change in its state.

\begin{figure}
\vspace{-0.9cm}
\centerline{\psfig{figure=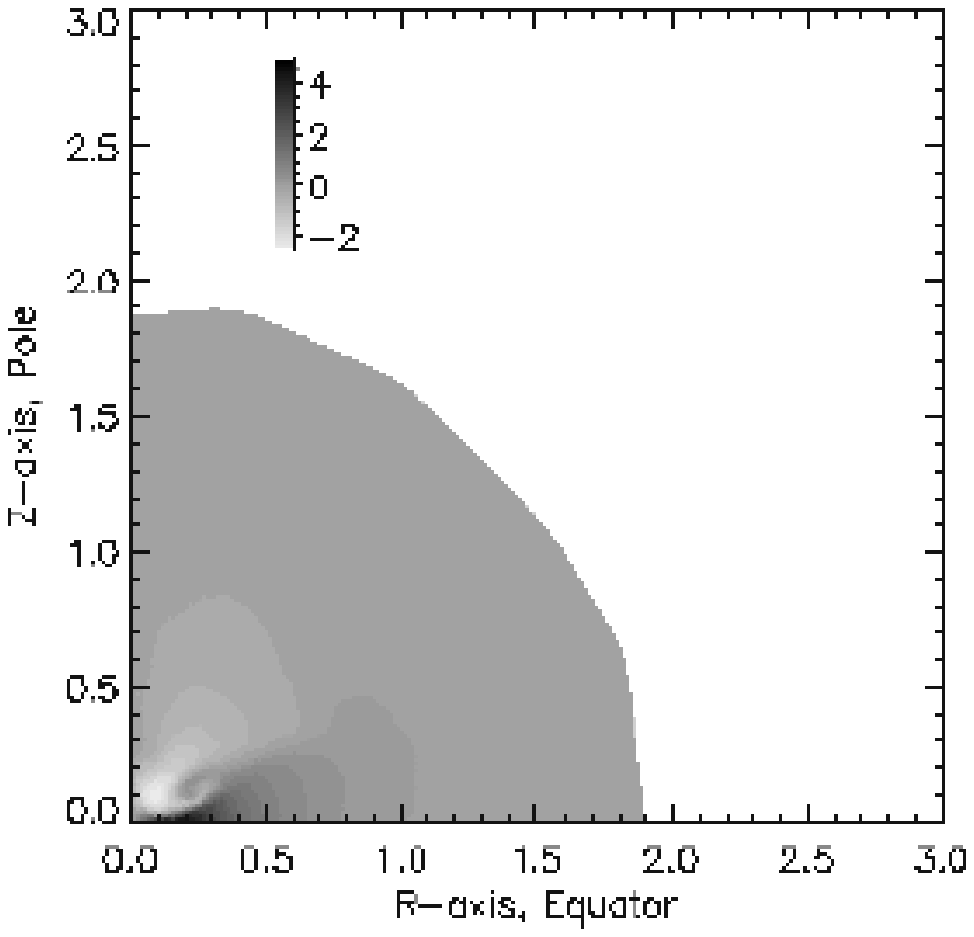,height=8.5cm,width=8.5cm}}
\vspace{-8.5cm}
\centerline{\psfig{figure=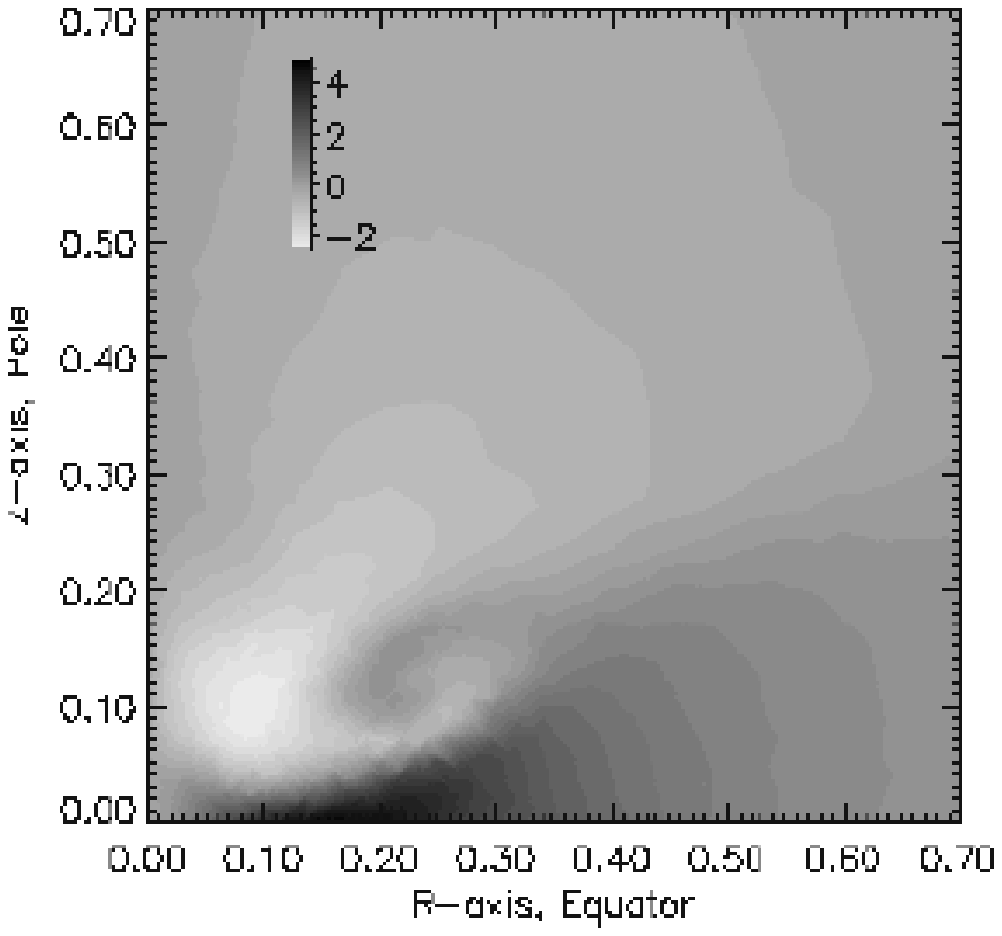,height=8.5cm,width=8.5cm}}
\caption{Toroidal magnetic field at $t=28.35$.}
\label{torf}
\end{figure}

\begin{figure}
\centerline{\psfig{figure=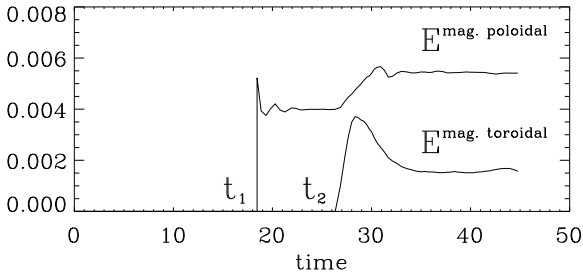,height=4.25cm,width=8.5cm}}
\caption{Time evolution of poloidal and toroidal magnetic energies.
$t_1$ is the moment of turning on the initial poloidal magnetic
field, $t_2$ is the moment of the beginning of the evolution of the
toroidal magnetic field.}
\label{magenerg}
\end{figure}

At $t_1$ we "turn on" a poloidal magnetic field calculated
as described above.
Simultaneously at $t_1$ we "switch off" the equation for the
evolution of the toroidal magnetic field $H_\varphi$ for a short
time (as discussed in the section "Initial magnetic field
configuration").
After a few weak oscillations around
equilibrium, the cloud comes to the steady state configuration
with a balanced magnetic force. These oscillations are clearly
seen in the plot for the time evolution of the poloidal magnetic
energy (Fig. \ref{magenerg}) This purely poloidal field becomes
balanced at $t_2=26.86$ . At this time we "switch on" the
equation for the toroidal component of magnetic field and let the
magnetic force lines twist due to differential rotation.

\begin{figure}
\centerline{\psfig{figure=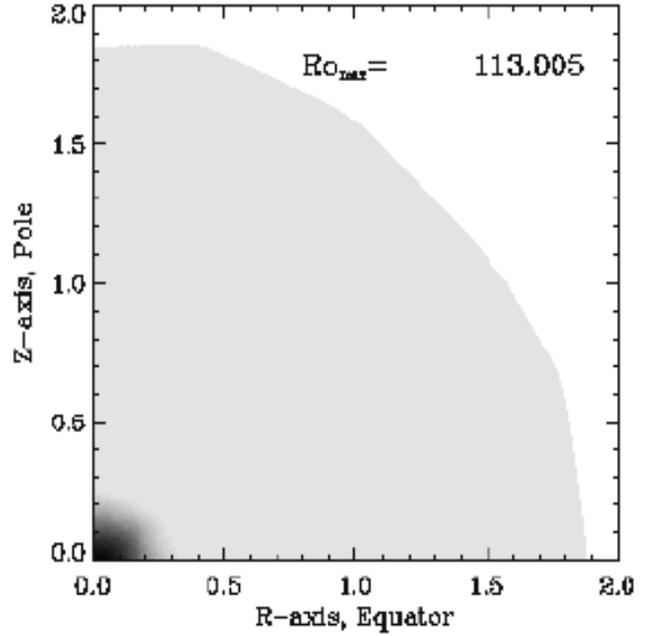,height=8.5cm,width=8.5cm}}
\caption{Density distribution at $t=28.0$, beginning the
evolution of the toroidal magnetic field. The black colour
corresponds to the maximal density.}
\label{densmax}
\end{figure}

Due to the differential rotation of the cloud and very
high (infinite) conductivity soon after $t=t_2$, the toroidal
component of magnetic field appears and grows with time. Taking
into account the quadrupole-like symmetry of the initial poloidal
magnetic fields the toroidal component appears and has 2 extremes
(Fig. \ref{torf}). A maximum at the equatorial plane and a minimum
at the periphery of the core of the cloud, close to the z-axis.
These extrema at the initial stage of the evolution of the
toroidal magnetic field correspond approximately to the extremes
of the following scalar product:  ${\bf H}\cdot{\rm grad}\frac
{V_\varphi}{r}$, because the cloud at the moment of "switching
on" the equation for toroidal magnetic field was in a steady
state condition and only this term in the equation for the
$H_\varphi$ determines the evolution of the toroidal field.

Toroidal magnetic energy grows almost linearly starting from
$t_2$ up to reaching its maximum value (Fig. \ref{magenerg})
after $\sim$ 2.5 rotations of the central core, which rotates
almost rigidly.
Increase of the toroidal magnetic field leads to
subtraction of angular momentum from the central parts of the
cloud and additional
contraction of the core of the cloud, meanwhile the envelope of the
cloud starts to blow up.
Contraction wave moving outwards appears at the periphery of the
core of the cloud. This wave propagates along a decreasing
density profile, increasing its amplitude and transforms to the
MHD shock wave.

\begin{figure}
\centerline{\psfig{figure=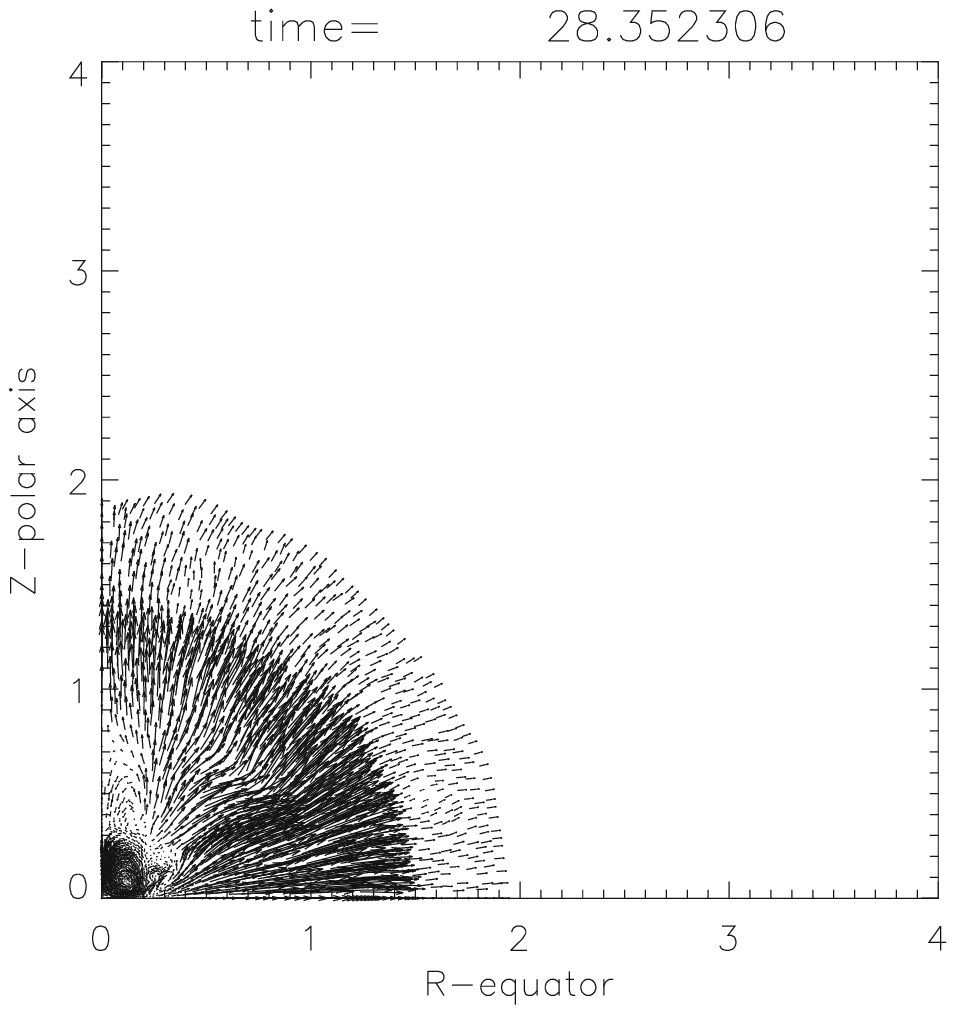,height=8.5cm,width=8.5cm}}
\vspace{-0.6cm}
\centerline{\psfig{figure=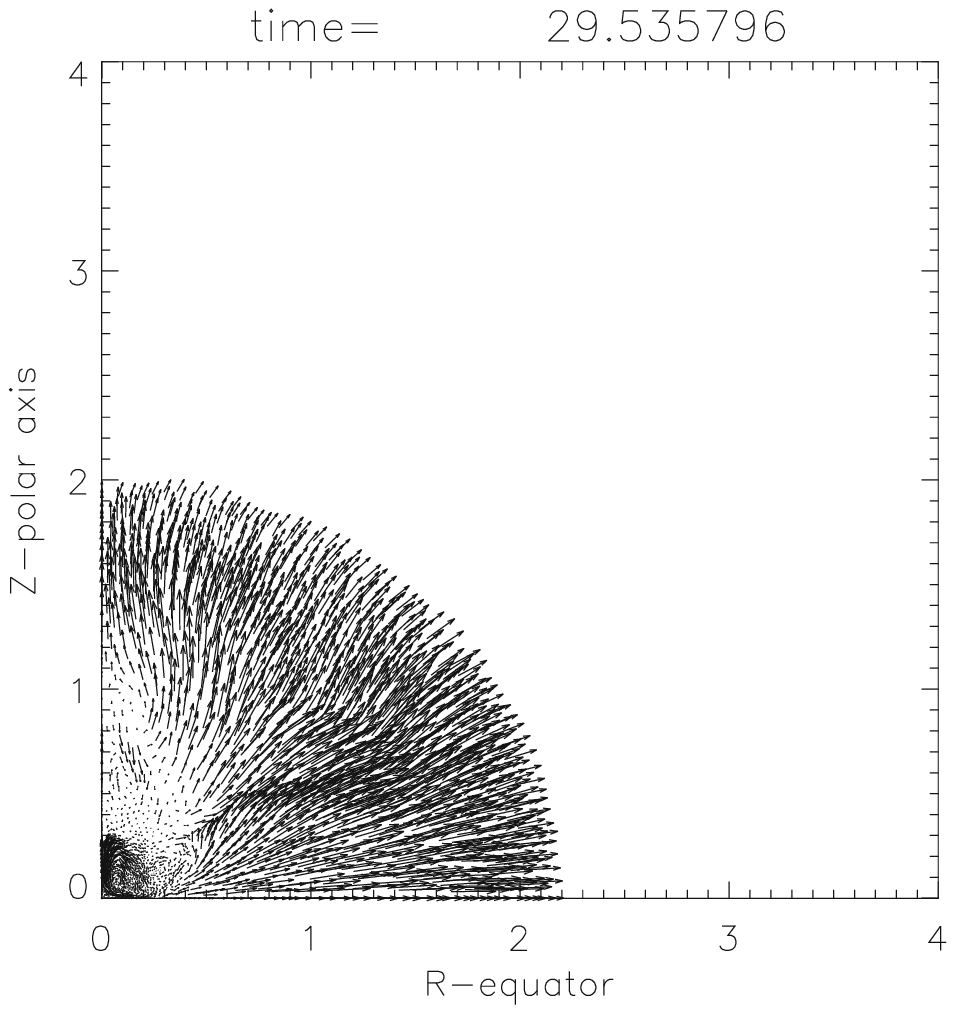,height=8.5cm,width=8.5cm}}
\vspace{-0.6cm}
\centerline{\psfig{figure=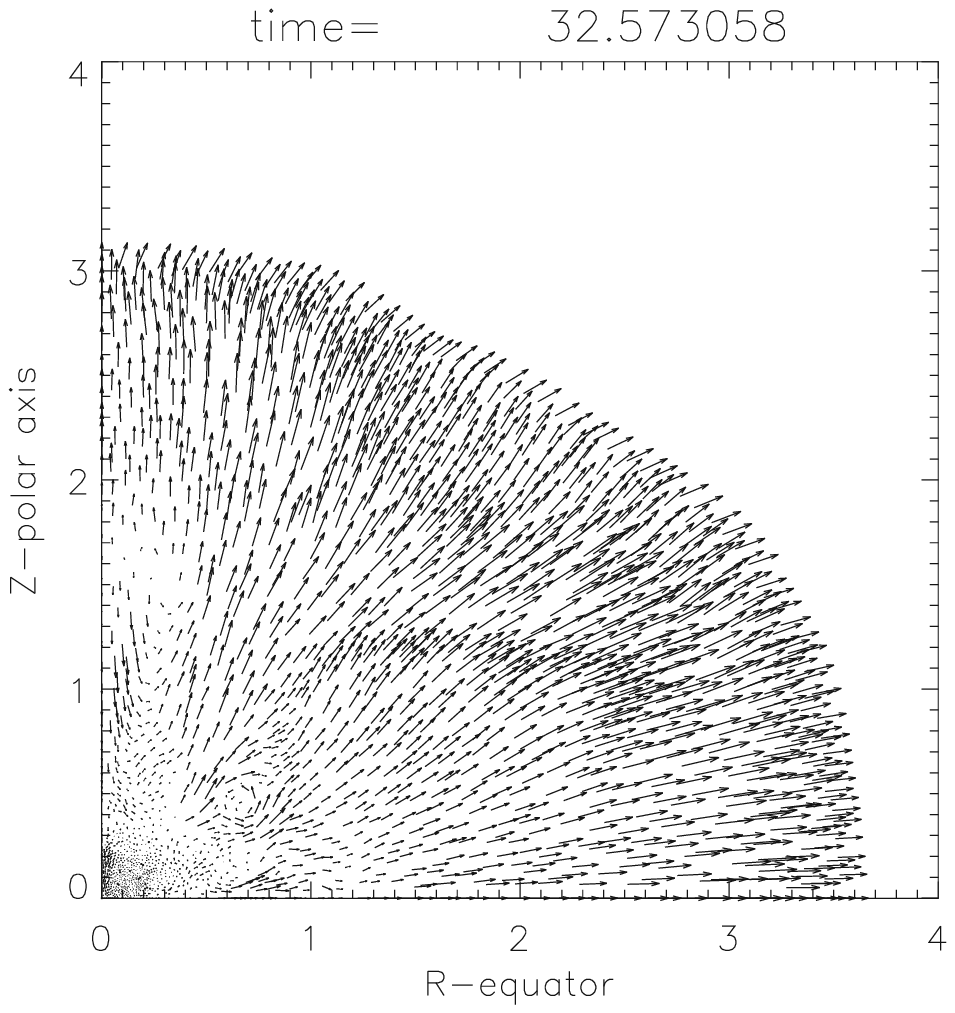,height=8.5cm,width=8.5cm}}
\vspace{-0.8cm}
\caption{Velocity field of the cloud.}
\vspace{-0.95cm}
\label{vfield}
\end{figure}

\begin{figure}
\centerline{\psfig{figure=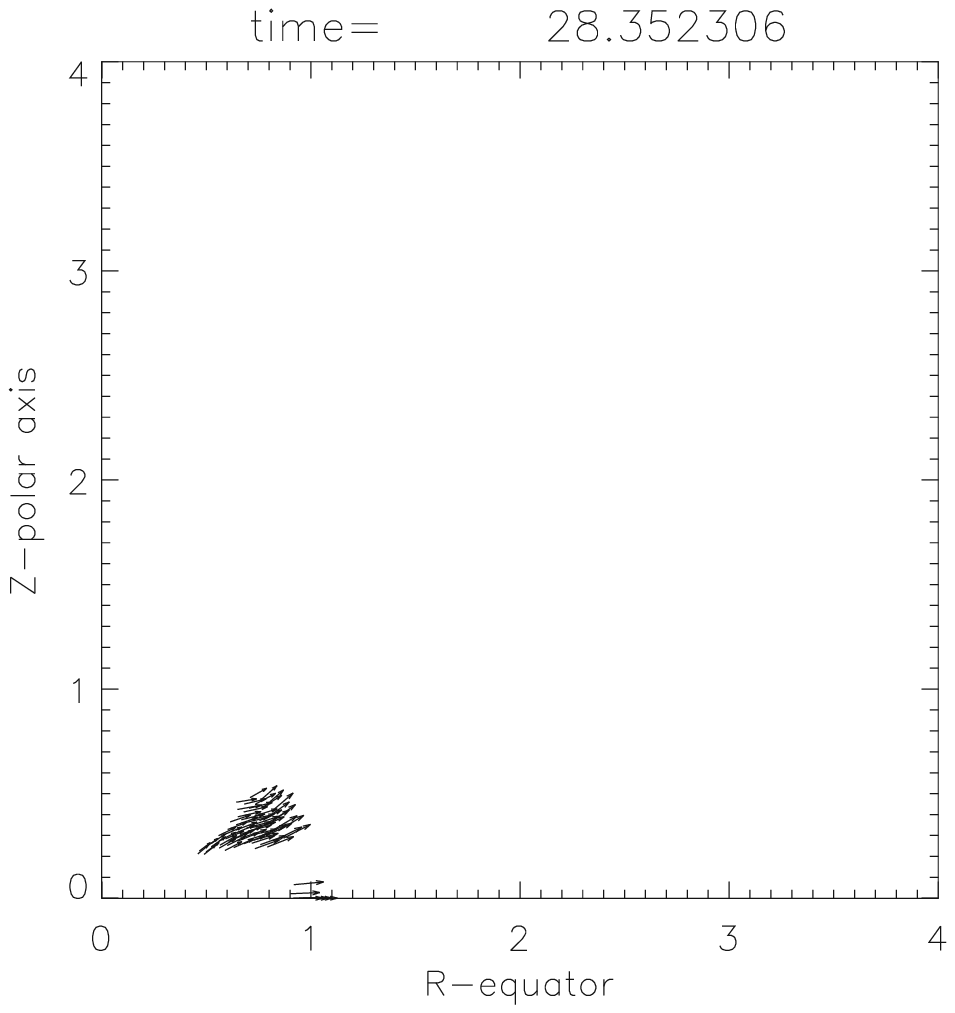,height=8.5cm,width=8.5cm}}
\vspace{-0.6cm}
\centerline{\psfig{figure=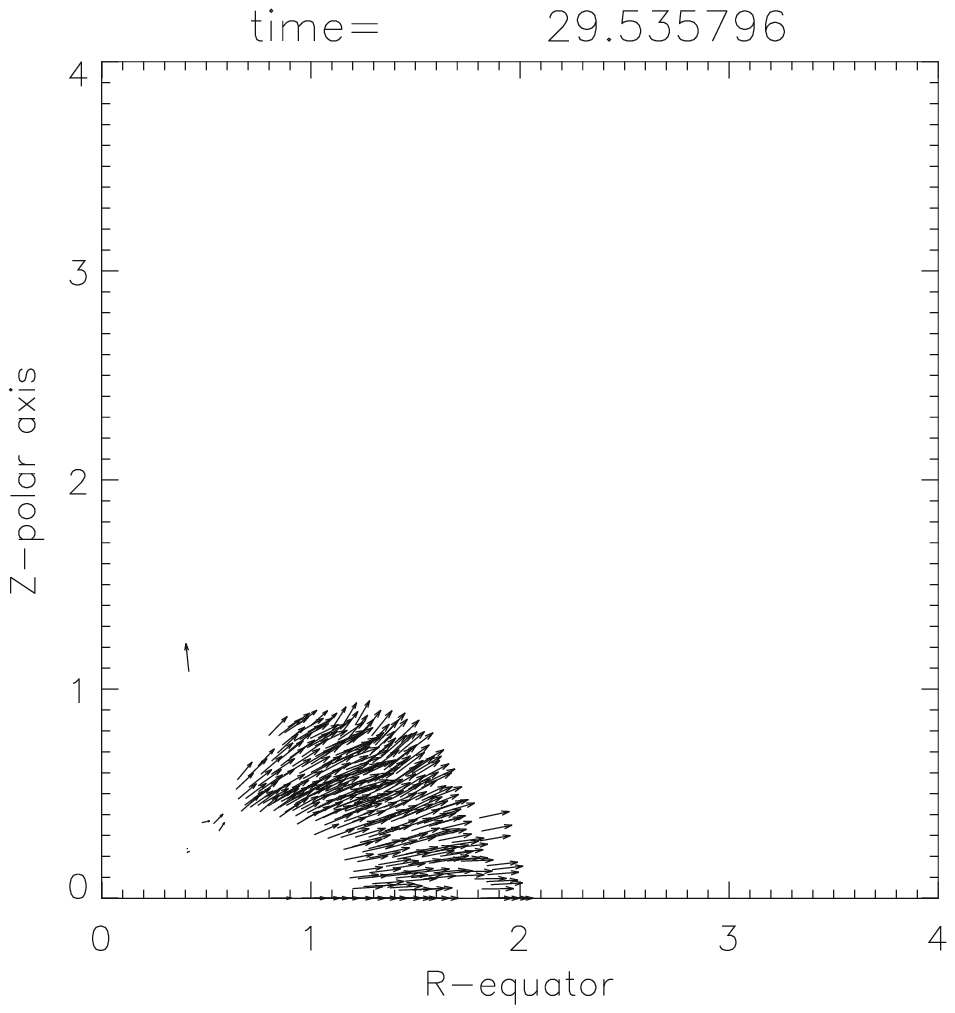,height=8.5cm,width=8.5cm}}
\vspace{-0.6cm}
\centerline{\psfig{figure=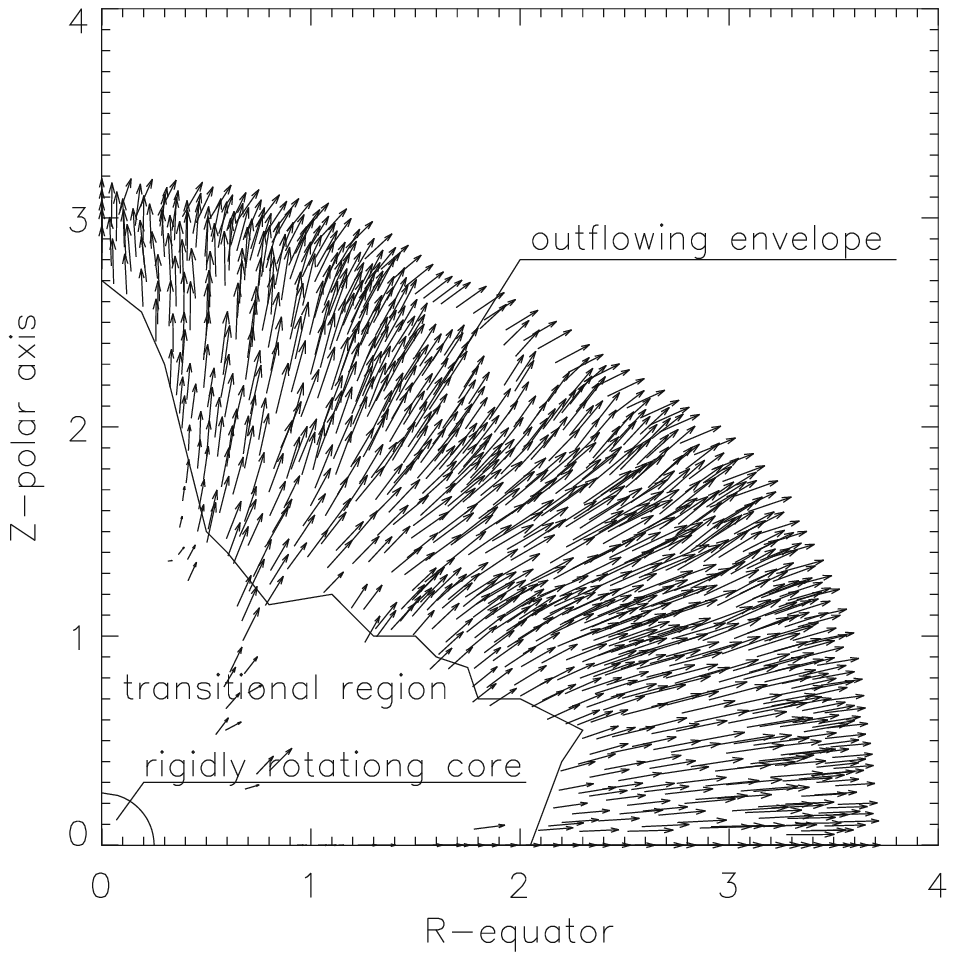,height=8.5cm,width=8.5cm}}
\vspace{-0.8cm}
\caption{Time evolution of the ejected part of the cloud.}
\vspace{-0.95cm}
\label{escape}
\end{figure}

At $t=28.0$ a small outer part of the envelope of the cloud
reaches the poloidal (radial) velocity ($v_r,\>v_z$), which
corresponds to a
kinetic energy larger then its potential energy and
can fly away to infinity.  At this time maximal density in
the center of the cloud is $\rho_{max} =113.0$
(Fig. \ref{densmax}). The amount of matter thrown away by the MHD
shock, grows with time. In figs.\ref{vfield},\ref{escape} one
can see the evolution of the velocity field and development of the
process of ejection of the matter of the envelope of the cloud
with time.

\begin{figure}
\centerline{\psfig{figure=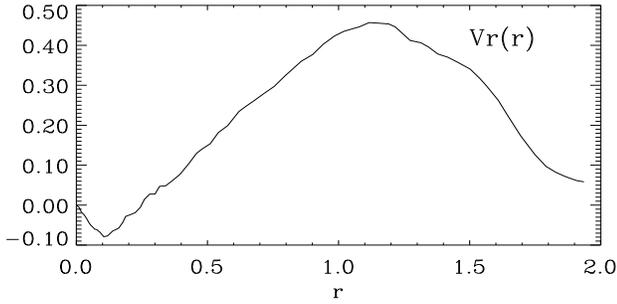,height=4.6cm,width=8.5cm}}
\caption{Radial velocity $v_r$
 distribution along the $r$ axis at $t_1=28.886$.}
\label{vrr}
\end{figure}

\begin{figure}
\centerline{\psfig{figure=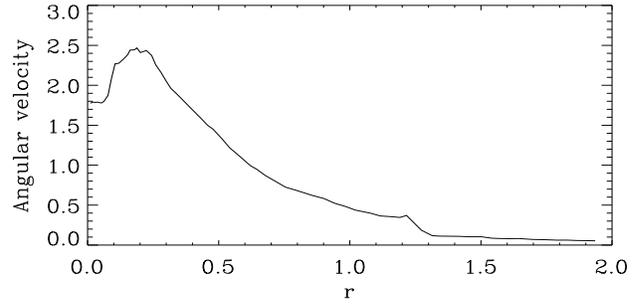,height=4.6cm,width=8.5cm}}
\caption{Angular velocity $V_\varphi/r$
 distribution along the $r$ axis at $t_1=28.886$.}
\label{omegar1}
\end{figure}

The growth of the toroidal magnetic field leads to the formation
of MHD shocks. To analyze the structure of the evolution of these
shocks we consider a flow picture at the equatorial plane
(along $r$-axis). The toroidal field at the $r$-axis,
has the maximum at the periphery of the core of the
cloud, approximately coinciding with the point where the radial
velocity $v_r$ changes its sign (Fig. \ref{vrr}).
At $r\approx 0.15$ (Fig. \ref{omegar1}) MHD
shock is formed, and it is a slow MHD shock, because its velocity is
lower than the local Alfvenic velocity. Angular velocity
$v_\varphi/r$ and $H_\varphi$ is decreasing, when passing through
this shock. The density $\rho$ and temperature $T$ are
increasing.

\begin{figure}
\centerline{\psfig{figure=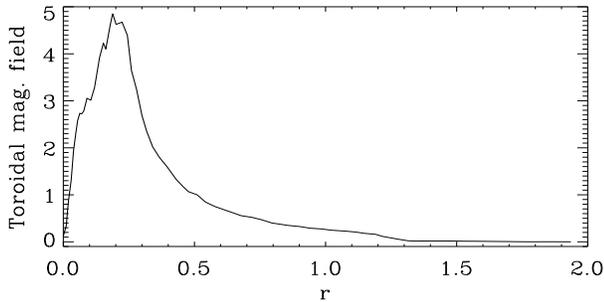,height=4.6cm,width=8.5cm}}
\caption{Toroidal magnetic field $H_\varphi$
 distribution along the $r$ axis at $t_1=28.886$.}
\label{hfr1}
\end{figure}

The matter which is on the right part of the maximum of
$H_\varphi$ is moving outwards. The MHD
shock, which is formed on the right side of the maximum of the
toroidal field at the equatorial plane is a fast MHD shock wave,
and its
velocity is bigger than fast magnetic sound speed before the
front of the shock. The velocity of the MHD shock wave is
smaller than the Alfvenic sound speed in the gas behind the shock,
and bigger than the slow magnetic sound speed after the shock
\cite{kulljub}.
The toroidal magnetic field (Fig. \ref{hfr1}) and angular velocity
(Fig. \ref{omegar1}) grow behind the shock.  Due to the
transformation of the rotational energy into the kinetic energy of
the radial motion, the angular velocity of the core of the cloud
decreases with time (Fig.\ref{omegar1}). At $t=28.886$ the shock
is at $r\approx 1.25$ (see for example Fig.\ref{omegar1}).

\begin{figure}
\centerline{\psfig{figure=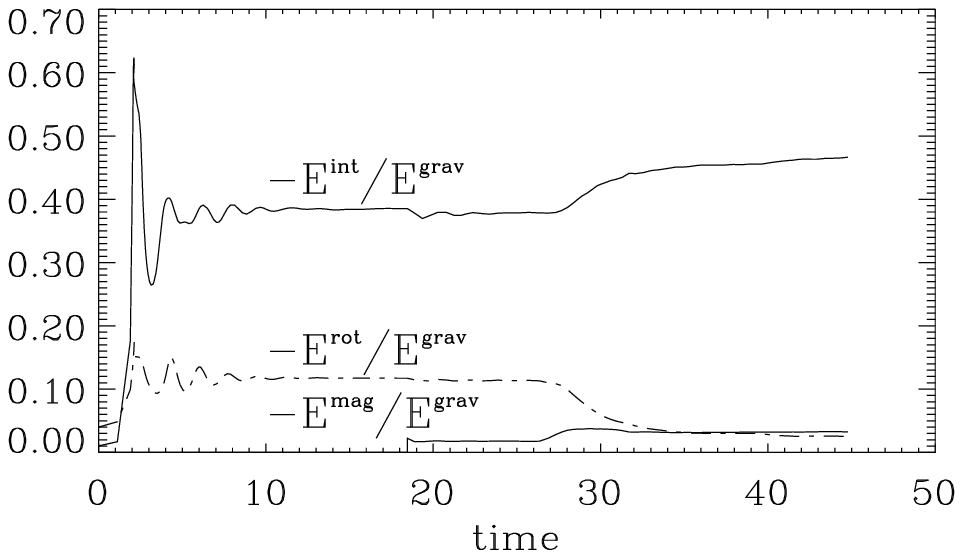,height=5.cm,width=8.5cm}}
\caption{Time evolution of parameters
$\xi,\>\beta_i,\>\beta_r.$ for $\xi(t_1)=10^{-2}$}
\label{albega-2}
\end{figure}

Similar slow MHD shock  has been found in 1D
simulations of magnetrotational supernova explosion made by
Ardeljan, et.al. (1979), while the formation of the fast MHD shock was
not presented.

The time evolution of the parameters
\begin{equation}
\beta_r(t) =\frac {E_{\rm rot}(t)}{\vert E_{\rm gr}(t)\vert },\>
\beta_i(t)  =\frac {E_{\rm in}(t)}{\vert E_{\rm gr}(t)\vert },\>
\xi(t) =\frac {E_{\rm mag}(t)}{\vert E_{\rm gr}(t)\vert },
\end{equation}

for initial $\xi(t_1)=10^{-2}$, is given in
Fig.(\ref{albega-2}).

\begin{figure}
\centerline{\psfig{figure=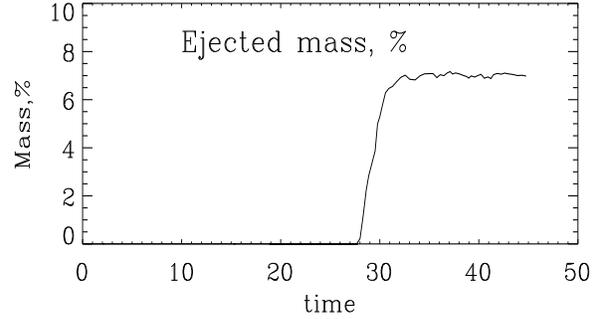,height=5.cm,width=8.5cm}}
\caption{Time evolution (in \%) of the ejected mass of the
envelope of the cloud.}
\label{uletmas}
\end{figure}

\begin{figure}
\centerline{\psfig{figure=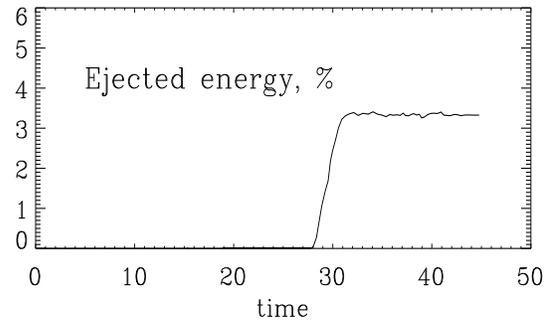,height=5.cm,width=8.5cm}}
\caption{Time evolution (in \% to the total initial energy of
the cloud) of the energy of the ejected matter of the
envelope of the cloud.}
\label{uletek}
\end{figure}

In Figs.~\ref{uletmas} and ~\ref{uletek}, time evolution is
presented of the
ejected mass of the envelope of the cloud, and the energy it
carries away. Just after the beginning of the
ejection the amount of the ejected mass grows approximately
linearly with time.
At the moment $t=33.57$, its growth stops and until the end
of the calculations these values (ejected mass and energy) do not
change significantly. The amount of the ejected matter after
this time is about 7\% and it carries approximately 3.3\% of the
total energy of the cloud. At the moment of the evolution of
the magnetorotational explosion the rotation of the cloud changes
essentially. The core of the cloud now rotates with the small
angular velocity $\omega_{core}\approx 0.85$.
The ejecting matter now looks like an
expanding shell.
At $t=28.0$ (the moment of the "switching on" of the evolution of
the toroidal magnetic field, the "inner" (core and part of the
envelope) 50\% of the mass of the cloud contained 26.8\% of
the angular momentum, at $t=33.57$ the "inner" 50\% of the mass
of the cloud contained 7.2\% of the angular momentum. At
$t=28.0$ the 50\% of the angular momentum are contained
in the outer part of the envelope of the cloud (31.3\% of the
mass of the cloud). At $t=33.57$ the 50\% of the angular momentum
is concentrated in the outer part of the envelope of the cloud
(9.2\% of the mass of the cloud). At the last plot of
Fig. \ref{escape}
the structure of the cloud at the advanced stage
of the magnetorotational
explosion is given. At this stage the cloud
consists of an outflowing envelope,
a transitional region and an almost rigidly rotating core.

In the paper by \cite{ardbkpop} it was found in 1D calculations
of the magnetorotational supernova explosion that during the
evolution of the toroidal component and angular momentum
transfer outwards, the central core of the cloud starts to rotate
in the opposite direction to the initial one. Such
magnetorotational oscillations have been investigated
analytically by \cite{bkpopsam}. With respect to the star
formation problems in slightly different formulation, these
oscillations  have been investigated in the papers by
\cite{mospal79}, \cite{mospal80}. In our simulations we do not
get the opposite rotation of the core of the cloud in the opposite
direction. However the core loses significant part of its angular
momentum due to magnetic breaking.

\section{Discussion}


The first 2D calculations of the magnetorotational explosion were
performed by Le Blank \& Wilson (1970), where matter was
expelled in the form of jets. Such geometry of the outburst was
the result of a specific rather unrealistic choice  of the
magnetic field configuration, which was produced by a current
ring, at an equator, out of a stellar centre, where the matter
density was an order of magnitude less than a central one.
The magnetic field of this ring had a zero radial component in the
equatorial plane, and a magnetic pressure gradient was formed in
the $z$-direction due to the differential rotation, which almost
repeated the magnetic pressure gradient  of the initial
configuration. Such magnetic field structure had led to the
matter stream pattern appearing preferentially along the symmetry
axis of the magnetic field.

In our calculations the magnetic field was created by the oppositely
directed current rings in the central region of the star. The
magnetic field had a quadrupole-like configuration with a
maximum value close to the centre of the star, and large radial
component at the equator.

In the papers devoted to the simulations of the star formation
problems (e.g. Ciolek \& Mouschovias, 1995; Tomisaka, 1998) it is
often supposed that the initial poloidal magnetic field has only an
$H_z$ component. It is often suggested for the collapse problem of
protostellar clouds. We should point out that the initial
magnetic field
of the configuration $H_{0z}\ne 0,\> H_{0r}=0$ or dipole-like
magnetic fields ($H_{0r}=0$ at the equatorial plane) can lead
after, the evolution of the toroidal component, to the jet-like
outflows directed presumably along the $z$-axis. However application
of the quadrupole-like initial poloidal field ($H_z=0$ at the
equatorial plane) can give radial ejections, as is seen in our
results.

The initial magnetic field configuration in the rotating gas
cloud as a progenitor of the star formation may be rather
complicated because of noncoincidence of magnetic and rotational
axes and possible action of the dynamo mechanism. Due to this fact
the prevailing of a quadrupole-like component in the initial model
it seems to be possible in some cases also in the cloud.

Observation of the solar corona during sunspot minimum has shown a
north-south asymmetry, interpreted by Osherovich et al. (1984) as
the
presence of a significant magnetic quadrupole. Due to more rapid
central increase of the quadrupole component ($\sim 1/r^4$) in
comparison with the dipole ($\sim 1/r^3$), we may expect that
quadrupole component may be larger than
the dipole one in the central part of the star.
Besides, we believe that the maximum of the magnetic
field strength can not be situated at such large distance from
the centre, as it was taken by Le Blank \& Wilson (1970).
Another 2D calculation, performed by Ohnishi (1983) for a radial
initial magnetic field gave an outburst mainly in the equatorial
plane, similar to our results.

The amount of energy of the matter ejected during the
magnetorotational explosion is sufficient for producing a
supernova explosion after formation of the rapidly rotating
neutron star with a differentially rotating envelope. Such calculations
are now in progress.


\appendix
\section{Numerical method }

We have used a numerical technique, based on a conservative
(in the absence of gravitation), implicit first order
of accuracy in space and time operator difference scheme on a
triangular grid with a grid reconstruction, developed and
described in  \cite{ardbook}, \cite{ardkos}. This numerical method
has been
used in the paper by \cite{ardmsgaa} for the investigation of
the collapse of the nonmagnetized rotating cloud.

For a numerical solution of the problem of the
magnetorotational explosion, we introduce a triangular grid,
covering the restricted domain in $r,\>z$ coordinates.

We suppose that components of the velocity vector {\bf u} and
gravitational potential $\Phi$ will be defined in all knots of
the grid. The density $\rho$, pressure $p$, components of
magnetic field vector {\bf H} will be defined in the cells and
in the boundary knots of the grid.

For the numerical simulation of the system of gravitational
MHD equations the following method has been used. Instead of
differential operators ($\rm div, \> grad, \> rot$) we introduce
their finite difference analogues.
On the base of such operators a completely conservative scheme
has been constructed. The scheme is implicit for all velocity
components $v_r,v_\varphi,v_z$ and for the toroidal component of
the magnetic field $H_\varphi$. The scheme is explicit for the
poloidal magnetic field $H_r,H_z$. The explicitness of the
scheme for $H_r,H_z$ does not introduce strong restriction on
the time step, because during the evolution of the magnetic
field its poloidal values do not change strongly,
while the toroidal component appears and increases
 significantly with time.
The scheme is explicit for the gravitational
potential, but it was shown in \cite{ardbook} that this
explicitness does not introduce significant restrictions on the
time step.


\subsection{Calculation of the initial and boundary
values of the magnetic field}

Initial values of the poloidal components of the magnetic field
$H_{r_0},\> H_{z_0}$ in the computational domain and its
boundary values at the outer boundary $H_{rq},\> H_{zq}$ are
calculated using the Bio-Savara law:

\begin{equation}
\label{biosavara}
{\bf H}(M_0)=\frac 1 c \int\limits_V^{}
\frac {{\bf J} \times {\bf R}_{MM_0}}{{R_{MM_0}}^3}dV_M,
\end{equation}
where ${\bf J}=j_r{\bf e}_r+j_\varphi{\bf e}_\varphi +
j_z{\bf e}_z$ - current density,
${\bf e}_r,{\bf e}_\varphi,{\bf e}_z$ -
unit vectors of cylindrical system of coordinates, $c$ - light
velocity, ${\bf R}_{MM_0}$ - radius vector connecting $M_0$ and
$M$ points, $R_{MM_0}$ - length of the vector ${\bf R}_{MM_0}$.

In the cylindrical coordinates we have:
\begin{equation}
\label{biosavararas}
{\bf H}(M_0)= {1
\over c} \int\limits_S rds
\left(\int\limits_0^{2\pi} \frac {{\bf J} \times {\bf
R}_{MM_0}}{{R_{MM_0}}^3}d\varphi\right).
\end{equation}
where $S$ is the computational domain in $r,z$ coordinates.

On the triangular grid the integral $\int\limits_S f(r,z) rds$ can
be approximated by the following sum:
\begin{equation}
\int\limits_S f(r,z) rds \approx
\sum_{\triangle \in \omega_i} r_is_i f(r_i,z_i)
\end{equation}
where $\omega_i$ is a set of triangular cells,
$r_i$ is $r$ coordinate of the center of the cell $i$,
$z_i$ is $z$ coordinate of the center of the cell $i$,
$s_i$ -- area of the
grid cell $i$.

For the integration of the vector-function in cylindrical
coordinates we use the following formula:
\begin{eqnarray}
\int\limits_{\varphi_1}^{\varphi_2}
\Big[f_r(\varphi){\bf e}_r+f_\varphi(\varphi){\bf e}_\varphi +
f_z(\varphi){\bf e}_z \Big]d\varphi  \nonumber
\end{eqnarray}
\begin{eqnarray}
={\bf e}_r(\varphi_2)\int\limits_{\varphi_1}^{\varphi_2}
\Big[f_r(\varphi) \cos(\varphi_2-\varphi)+ f_\varphi(\varphi)
\sin(\varphi_2-\varphi)\Big] d\varphi  \nonumber
\end{eqnarray}
\begin{eqnarray}
+{\bf e}_\varphi(\varphi_2)\int\limits_{\varphi_1}^{\varphi_2}
\Big[f_\varphi(\varphi) \cos(\varphi_2-\varphi)
-f_r(\varphi) \sin(\varphi_2-\varphi)\Big]
d\varphi  \nonumber
\end{eqnarray}
\begin{eqnarray}
+{\bf e}_z(\varphi_2)\int\limits_{\varphi_1}^{\varphi_2}
f_z(\varphi) d\varphi. \label{integrcyl}
\end{eqnarray}
Using~(\ref{integrcyl}) for~(\ref{biosavararas}), after
transformations we get:

\begin{eqnarray}
H_r(r,z)=\sum\limits_{\triangle_i \in \omega_i}^{}
r_i s_i j_{\varphi i}
\Bigg\{
\nonumber
\end{eqnarray}
\begin{eqnarray}
  (z-z_i)\frac {4}{a_1^3}
  \Bigg[
    \frac {2-k_1^2}{k_1^2} \Pi\big( -k_1^2,k_1,\frac {\pi}{2} \big)-
    \frac {2}{k_1^2} F \big( k_1, \frac {\pi}{2} \big)
  \Bigg]
\nonumber
\end{eqnarray}
\begin{eqnarray}
    +(z+z_i)\frac {4}{a_2^3}
  \Bigg[
    \frac {2-k_2^2}{k_2^2} \Pi \big( -k_2^2,k_2,\frac {\pi}{2} \big)-
    \frac {2}{k_2^2} F\big( k_2, \frac {\pi}{2} \big)
  \Bigg]
\Bigg\},
\nonumber
\end{eqnarray}
\begin{eqnarray}
\label{biosavr}
H_\varphi(r,z)=0,
\end{eqnarray}
\begin{eqnarray}
H_z(r,z)=\sum\limits_{\triangle_i \in \omega_i}^{}
r_i s_i j_{\varphi i}
\Bigg\{ r_i\frac 4 {a_1^3} \Pi \big(-k_1^2,k_1,\frac \pi 2 \big)
\nonumber
\end{eqnarray}
\begin{eqnarray}
+r \frac 4 {a_1^3}
\Bigg[ \frac {2-k_1^2}{k_1^2}\Pi \big(-k_1^2,k_1,\frac \pi 2 \big)
 - \frac 2 {k_1^2} F\big(k_1, \frac \pi 2\big)
\Bigg]
\nonumber
\end{eqnarray}
\begin{eqnarray}
+r_i \frac 4 {a_2^3} \Pi\big(-k_2^2,k_2,\frac \pi 2 \big)
\nonumber
\end{eqnarray}
\begin{eqnarray}
-r \frac 4 {a_2^3}
\Bigg[ \frac {2-k_2^2}{k_2^2}\Pi \big(-k_2^2,k_2,\frac \pi 2 \big)
 - \frac 2 {k_2^2} F\big(k_2, \frac \pi 2\big)
\Bigg]
\Bigg\}.
\nonumber
\end{eqnarray}

\noindent
Here
\begin{eqnarray}
{k_1}^2={4r_ir \over {a_1}^2}, \> \> {a_1}^2=(r_i+r)^2+(z-z_i)^2,
\nonumber\\
{k_2}^2={4r_ir \over {a_2}^2}, \> \> {a_2}^2=(r_i+r)^2+(z+z_i)^2,
\nonumber
\end{eqnarray}

\noindent
$j_{\varphi i}$ is the toroidal component of the current
in the center of the cell $i$,

\noindent
$F(k,{\pi \over 2})=\int\limits_0^{\pi/2}{{dt}\over
{\sqrt(1-k^2{\rm sin}^2t)}}$ --  elliptic integral of the first kind,

\noindent
$\Pi(-k^2,k,{\pi \over 2})=\int\limits_0^{\pi/2}{{dt}\over
{(1-k^2{\rm sin}^2t)^{3/2}}}$ -- elliptic integral of the third kind.

Taking into account, that ${\bf J}={c \over 4\pi} {\rm rot} {\bf
H}$, and using~(\ref{biosavr}) we get boundary values of the poloidal
magnetic field at the outer boundary.

Approximation error for boundary values of the poloidal
components of the magnetic field in the formulae~(\ref{biosavr}) is
$O(h)$. However using these formulae requires $O(N\sqrt{N})$
operations (where $N$ is the total number of grid knots) and takes a
lot of CPU time. During calculation the values of the poloidal
components of magnetic field at the outer boundary change
weakly and to decrease CPU time we used formula~(\ref{biosavr})
every tenth time step. At all other time steps we
extrapolate poloidal fields from the vicinity of the
outer boundary to the outer boundary.

\subsection{Testing}

Testing of the  numerical technique,
applied in this paper,
without magnetic field has been described in detail by
\cite{ardmsgaa}.

\noindent
In MHD case we add the following tests:
\begin{enumerate}
\item We calculated a spherically symmetrical collapse of
      a nonrotating
      gas cloud without a magnetic field until the cloud
      comes to the
      steady state condition.  After that we "turned on"
      force-free and divergency-free magnetic field $H_r=0,\>
      H_\varphi=0,\> H_z={\rm const}$.  This magnetic field does
      not change the condition of the cloud after hundreds of
      numerical time steps.
\item {\it The piston problem.}\footnote{The numerical data,
      used in this test are taken from \cite{sampop}} Consider
      the flat piston, which is pushed along the $r$-axis (our
      domain for this test was a rectangle with the following
      coordinates: ($r_0=1000,\> z_0=0;\> r_1=1001,\> z_1=1$),
      in this case we can consider $r,\>z$ as Cartesian
      coordinates with high accuracy)
      into the gas with following characteristics: infinite
      conductivity ($\sigma=\infty$), equation of state:

      $$
      p=\Re \rho T,\> \varepsilon =\Re T / (\gamma -1),
      \> \Re=1,\> \gamma=5/3.
      $$
      At $t=0$ the gas has the following parameters:
      $$
      {u_r}_0=0,\> {u_z}_0=0,\> \rho_0=1,\> T_0=0,
      $$
      $$
      {H_r}_0=2.507,\> {H_z}_0=1.401.
      $$
      The piston has a velocity:
      $$
      v_{r\>{\rm piston}}=0.25, \> v_{z\> {\rm piston}}=-0.2735.
      $$
      The values of the parameters of the gas after the shock wave
      front (defined by the Hugoniot formulae) are the following:
      $$
      u_r=0.25,\> u_z= -0.2735,\> T=0.0117,\> \rho = 1.33,
      $$
      $$
      H_r=2.507,\> H_z=2.802.
      $$

\end{enumerate}

The discrepancy in
the analytical and numerical values after the MHD shock was $<1\%$
for the magnetic field and $<0.5\%$ for the density. The same
test has been made for the shock, moving along $z$-axis with similar
discrepancy.

\begin{acknowledgements}
S.G.M. and G.S.B.-K. are grateful for
the partial support to RFBR by grant 99-02-18180, grant of the
Russian ministry of science "Astronomy program"
No. 1.2.6.5.
N.V.A. is thankful for the partial support to the Russian
ministry of education, program "Universities of Russia -
fundamental researches".

\end{acknowledgements}

\end{document}